# Applying Text Mining to Protest Stories as Voice against Media Censorship

**Tahsin Mayeesha**
North South University
Dhaka, Bangladesh

**Zareen Tasneem**
North South University
Dhaka, Bangladesh

**Jasmine Jones**
Univesity of Minnesota
author3@anotherco.com
jazzij@umn.edu

**Nova Ahmed**
North South University
Dhaka, Bangladesh
nova@northsouth.edu

## Abstract
Data driven activism attempts to collect, analyze and visualize data to foster social change. However, during media censorship it is often impossible to collect such data. Here we demonstrate that data from personal stories can also help us to gain insights about protests and activism which can work as a voice for the activists.

## ACM Classification Keywords
H.5.m. Information interfaces and presentation: Miscellaneous

## Introduction
Many social movements like "Occupy Wall Street"[1] or "Arab Spring"[2] has been quantified and modeled extensively using data from social media like Twitter. Despite lacking in social media data due to censorship, here we demonstrate that analyzing social movements by text analysis of personal stories can also help us to learn the emotional effects and entities involved in a social movement and can act as a voice for the activists. We use the data from a recent student driven protest in Bangladesh for road safety for this purpose.

The protest started from 29th July, 2018 but continued for about a month. A public bus competing with another bus lost control and ran over two school students [3, 4]. Students immediately initiated a movement seeking justice and safe road conditions. However, students were attacked by police and unidentified goons many times. Internet and social media based posts regarding the protest were restricted.

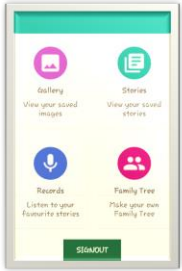

(a)

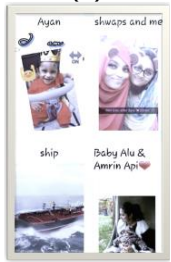

(b)

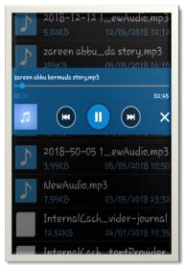

(c)

Figure 1 (a).Front View (b) Photo Sharing Option (c) Voice Recording Option of Golpokotha

We have extracted locations and organizations involved in the protest through network analysis and performed emotion mining on the collected story text to understand the unspoken emotions of the storytellers in depth. We aim to share our insights with the protesters and broader media to give visibility to the protest as well as help the protesters to make more sense of the events. With our work we aim to provide the vulnerable protesters who were oppressed and denied of freedom of speech during a legitimate protest for a basic need like safe roads a voice with data.

## Data Overview

We collect data through recordings of informal stories with open ended queries. Storytellers were recruited through established connections without phone calls or emails to protect privacy from two different cities, Dhaka and Chittagong. They had unique positions during the incident (as a student present in the protest, passerby, as a teacher or a family member). They were asked queries on what they thought happened during the protest and how they felt about it. Features like age, profession, story duration, date and text from transcribed stories were collected.

## Story Collection

A story collection tool named Golpo Kotha developed by North South University HCI lab was used that acts as a digital repository of family stories. It is able to preserve stories, photos connected to stories and family connection information. The application interface is shown in Figure 1. We have used the voice based story recording part for the current story collection. The listeners were in charge of recording the stories.

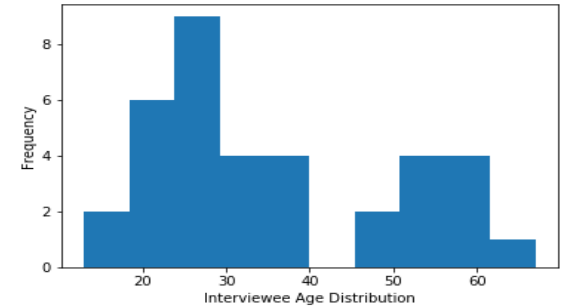

Figure 1 : Participant age distribution. Min age = 13 and Max age = 67.

## Story Analysis

### Participant Demographic and Stories:

36 sample stories were collected from 23 unique participants with varying age, gender and profession. Majority of the participants were students. Story duration distribution graph is given in figure 3. Median story is 1 minute long.

### Finding locations by Network Analysis:

In order to find the unique locations relevant to the protest, we have used network analysis and visualization techniques.

From the incidents described in the 36 sample stories, 19 unique locations were extracted. If two locations occur in the same story, we consider an edge between them to build a location co-occurrence network. The resulting graph has 19 nodes and 21 edges with an average degree of 2.21. Top 5 locations are given in table 1.

As can be seen in Figure 4, from the location network two different clusters of nodes are visible around Dhandmondi and Bashundhara, which corresponds to two main locations of protest. Other organizations like Apollo, labaid(hospitals who took care of the wounded) and NSU, IUB, East west University are also present.

have used emotion mining. IBM-Watson provides machine learning open source tools for such tasks.

Using IBM-Watson Tone Analyzer[6] for each story, extracted giving emotion and conversational style related information for each story and each of the sentence in the stories. Multiple tones such as anger and fear can be present in one story. 70 tones were collected from 36 unique stories. Anger, sadness and fear are the dominant emotional tone while analytical and confident were the conversational style tones.

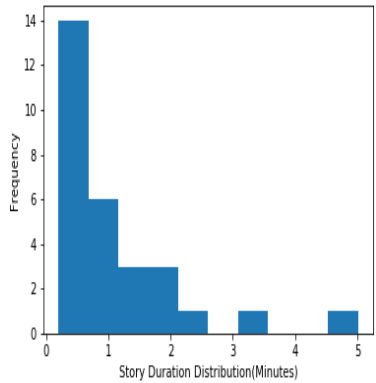

Figure 2 : Story Duration Distribution in Minutes. Longest Story is 5 minutes.

**Top 5 Locations by Degree Centrality**

| Location | Degree |
| --- | --- |
| Dhanmondi | 0.4444 |
| Apollo Hospital | 0.2777 |
| East West U | 0.2222 |
| Bashundhara | 0.2222 |
| NSU | 0.1667 |

Table 1: Degree Centrality for a node v is the fraction of nodes v is connected to. Degree centrality values are normalized by the maximum possible degree n-1 in a simple graph of n nodes.

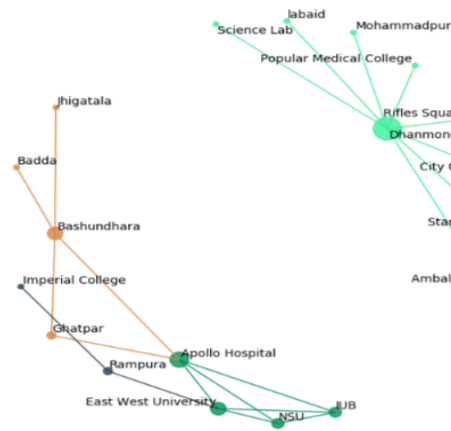

Figure 3 : Location Co-occurrence Network with Community Detection. Node colors are assigned randomly using Louvain method for community detection. [5]

**Mining Emotions From Stories:**
To understand the emotions reflected in the stories, we

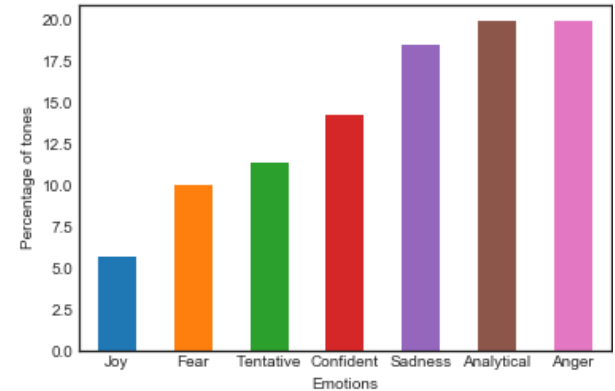

Figure 4 : Emotions Extracted from IBM Tone Analyzer.

## Text Visualization

Text visualization is used for showing the most frequently occurring words from the stories. Top 20 most common occurring words are presented in Figure 6. To visualize the text more generally we have used wordcloud.

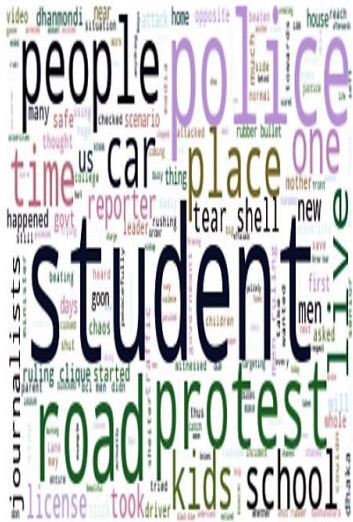

Figure 7 : Word Cloud from top 200 words

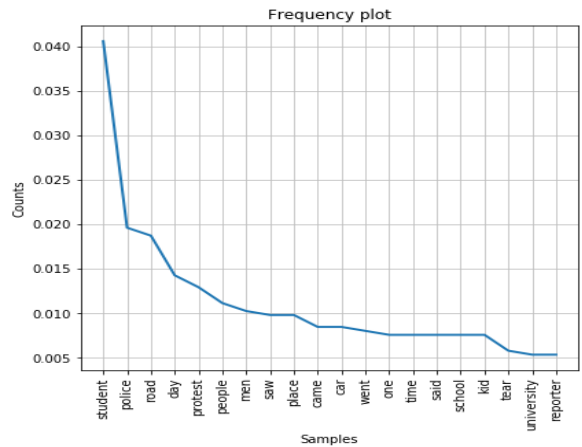

Figure 5 : Frequency plot of top 20 words

Incidentally some people have also shown joy for students work during the protest as they believed it will bring a better future. The tones are shown in Figure 5.

## Conclusion

With a very small dataset we have been able to determine relevant locations and emotions expressed in the stories. In future we plan to collect data from newspapers and compare with direct interviews and search for patterns using unsupervised machine learning methods like document clustering.

To conclude, text analysis techniques applied on activism narratives help us to summarize the events without looking at each of the stories manually and helps to find the hidden tones described in the narratives. Activists can also use the insights gained in their future work as well as share the insights with outside media to communicate with the rest of the world.

With our work and by submitting to CSCW we hope to connect the voice of the student protesters with the broader community and stand in solidarity with them publicly like UN[7] and European Union.[8]